\documentclass[final,5p,times,twocolumn]{elsarticle}

\usepackage{amssymb}
\usepackage{amsmath}
\usepackage{xcolor}
\usepackage{siunitx}
\usepackage{hyperref}
\usepackage{subcaption}

\journal{Nuclear Instruments and Methods in Research A}

\begin{document}

\begin{frontmatter}

\title{Beam Tests of SNSPDs with 120 GeV Protons}


\author[PHY]{Sangbaek Lee}
\author[PHY]{Tomas Polakovic\fnref{fn1}}
\fntext[fn1]{These authors contributed equally to this work.}
\author[PHY]{Whitney Armstrong\fnref{fn1}\corref{cor1}}
\cortext[cor1]{Corresponding author}
\ead{warmstrong@anl.gov}
\author[CNM,MSD]{Alan Dibos}
\author[MSD,NIU]{Timothy Draher}
\author[FTBF]{Nathaniel Pastika}
\author[PHY]{Zein-Eddine Meziani}
\author[PHY,MSD]{Valentine Novosad}

\affiliation[PHY]{organization={Physics Division, Argonne National Laboratory},
            city={Lemont},
            state={IL 60439},
            country={U.S.A.}}
\affiliation[CNM]{organization={Center for Nanoscale Materials, Argonne National Laboratory},
            city={Lemont},
            state={IL 60439},
            country={U.S.A.}}
\affiliation[MSD]{organization={Materials Science Division, Argonne National Laboratory},
            city={Lemont},
            state={IL 60439},
            country={U.S.A.}}
\affiliation[NIU]{organization = {Department of Physics, Northern Illinois University},
            city = {Dekalb},
            state={IL 60115},
            country={U.S.A.}}
\affiliation[FTBF]{organization = {Fermi National Accelerator Laboratory},
            city = {Batavia},
            state = {IL 60510},
            country = {U.S.A.}}
\begin{abstract}
We report the test results for a \SI{120}{GeV} proton beam incident on 
superconducting nanowire particle detectors of various wire sizes. NbN devices 
with the same sensitive area were fabricated with different wire widths and 
tested at a temperature of \SI{2.8}{K}. The relative detection efficiency 
was extracted from bias current scans for each device. The results show that the wire width is a critical factor in determining the detection efficiency and larger wire widths than \SI{400}{\nm} leads to inefficiencies at low bias currents. These results are particularly relevant for novel applications at 
accelerator facilities, such as the Electron-Ion Collider, where cryogenic 
cooling is readily available.
\end{abstract}




\end{frontmatter}

\section{Introduction}
\label{sec:intro}

Superconducting Nanowire Single Photon Detectors (SNSPDs) have been successfully deployed as optical and infrared photon sensors with precise spatial and timing resolutions \cite{doi:10.1063/1.1388868, Natarajan_2012}.
SNSPD operating principles suggest that the energy deposited from charged 
particles similarly induce hot spot formation and the SNSPD can efficiently function as a precise particle sensor \cite{SUZUKI20082001, Sclafani_2012, Cristiano_2015, doi:10.1063/1.3506692, doi:10.1063/1.4740074, nano10061198}. It was recently proven that it performs well under a high magnetic field \cite{Polakovic:2019wrh}.
These characteristics align well with different particle detection applications and, when combined with the low operating temperatures offer new opportunities. A good example is their possible use in the Electron-Ion Collider (EIC)'s far-forward particle detection, which is essential to its science mission \cite{AbdulKhalek:2021gbh}. Reconstruction of these particles requires a detector with fast-timing, O(\SI{100}{\um}) pixel size. Geometrical acceptance of semi-conductor detectors satisfying these requirements is limited by the beamline magnets, which can be mitigated by operating SNSPDs within the frigid bore of superconducting magnets.

The allowed proton beam energy configurations at the EIC are 41, 100, \SI{275}{GeV} \cite{AbdulKhalek:2021gbh}. At the far-forward region, the scattered proton is expected to have a similar energy to the beam protons with low transverse momenta. Detecting protons with these sensors in this energy range has never been reported before. In this paper, we report on the results of a direct detection of protons at the FermiLab Test Beam Facility (FTBF) using this detector for the first time. We first introduce the experimental setup in Section \ref{sec:experiment}, followed by the presentation of our test results and a discussion of potential future tests to address the limitations of this study in Section \ref{sec:results}. Finally, we summarize the impact of our findings and outline a future outlook in Section \ref{sec:summary}.

\section{Experimental Setup}
\label{sec:experiment}

\subsection{Device Fabrication}

The sensors were fabricated out of \SI{12}{\nm} Niobium Nitride (NbN) thin film deposited by ion beam assisted sputtering~\cite{ionBeamNbN} on top of a \SI{300}{\um} intrinsic silicon substrate. Thin films were patterned into nanowires using electron beam lithography, followed by ICP reactive ion etching in 20:25 mixture of CHF$_3$:SF$_6$ plasma. A utility layer of Ti(\SI{10}{\nm})/Au(\SI{150}{\nm}) for contact pads and alignment/chip dicing marks was added by optical maskless lithography and lift-off process. The superconducting critical temperature of a fully finished sensor device was approximately \SI{7}{K}.

\begin{figure}[!htb]
  \centering
  \includegraphics[width=0.25\textwidth]{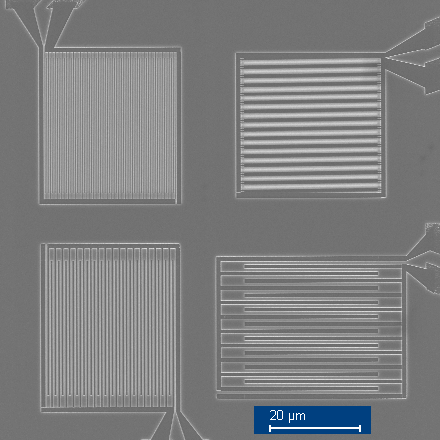}
  \caption{\label{fig:nanowires} SEM image of a typical test sensor array that has four different width configuration: \SI{100}{\nm} (top left), \SI{200}{\nm} (bottom left), \SI{400}{\nm} (top right), and \SI{800}{\nm}. }
\end{figure}

\begin{figure*}[!htb]
  \centering
  \includegraphics[width=0.8\textwidth]{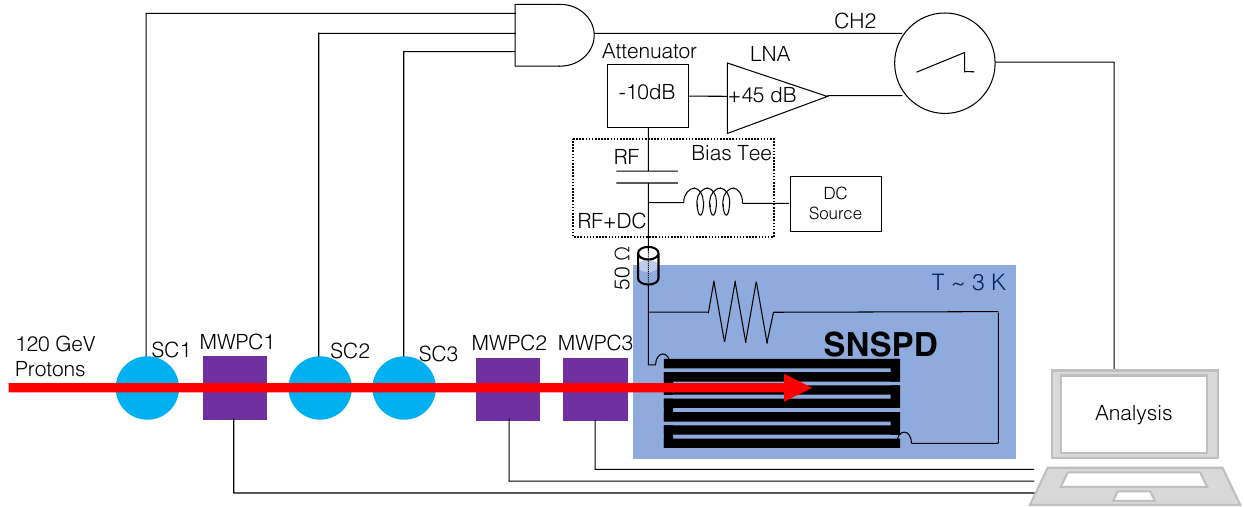}
  \caption{\label{fig:exp_schematics} A block diagram describing the experimental setup. The SNSPD detector was biased at variable current, whose amplified signal was recorded by the oscilloscope (CH1) and used as the self-trigger. The digital signals from the scintillation counters SC1, SC2, and SC3 were concurrently recorded at the oscilloscope (CH2). The signals from the MWPCs were collected separately and could be accessed via the local network.}
\end{figure*}

A typical single test chip (size \SI{8}{\mm}$\times$\SI{8}{\mm}) had 8 nanowire devices, each covering an active area of \SI{30}{\um}$\times$\SI{30}{\um} with a geometric filling factor of 1/2 (Fig.~\ref{fig:nanowires}). Sensors with wires widths of \SI{300}{\nm}, \SI{400}{\nm}, \SI{600}{\nm}, and \SI{800}{\nm} were tested. As these sensors are not designed to detect light, mostly due to their large nanowire width, an optical characterization was done only on a related device with width of \SI{150}{\nm}. The total efficiency of the test structure was determined to be approximatelly 3\% by counting from an attenuated calibrated \SI{405}{nm} laser, which is in the expected range of NbN-based SNSPDs without anti-reflective coating or optical cavities that are necessary to achieve near 100\% efficiency~\cite{li2018improving}. Readers interested in a more thorough optical characterization of devices fabricated using the same process are referred to work of reference~\cite{Polakovic:2019wrh}.

\subsection{Experimental Setup}
The chips were mounted and wire bonded to a printed circuit board (PCB). The PCB was mounted to a copper bracket coupled to the cold finger of the optical, two-stage GM cryo-cooler based cryostat (Janis SHI-4). During the test, the board and the sensor were held at a temperature of \SI{2.82}{K}. 
A chiller was installed inside the beam enclosure providing closed loop water-cooling for the helium compressor.
The measurements were performed at the FTBF using a 120 GeV proton beam in early 2023. At the Meson Test Section 6.2 (MT6.2), about 10$^6$ of \SI{120}{GeV} protons were delivered to the enclosure every minute with 4.2 s spill length on request. The protons were incident on the detectors as shown in Fig.~\ref{fig:exp_schematics}. 
The optical cryostat was modified with 1/16 inch Aluminum windows to minimize material thickness in the beam-line. 
Room temperature electronics placed in the beam enclosure near the cryostat include a biasing power supply, an attenuator and low noise amplifiers (LNAs).
The nanowire devices were current biased with a programmable constant current sources based on the LTC1427 current output DAC.
The waveforms were recorded on oscilloscopes via a coax cable patch panel connecting the MT6.2 beam enclosure and the electronics room, where the oscilloscopes and other trigger logic are located. 
The readout was triggered by the nanowire signal, and in addition to the nanowire  waveforms, the three-fold SC coincidence signal was also recorded as shown in Fig.~\ref{fig:protonSignal}.


Multi-wire Proportional Chambers (MWPCs) served to reference the lateral beam position and size with 1/$\sqrt{12}$ mm resolutions in both horizontal and vertical directions \cite{Fenker:1983km, LArIAT:2019kzd}. The optimal beam position was determined by scanning the dipole magnet currents that are nearly linear to the beam lateral coordinates in order to maximize the detection efficiency (Section \ref{subsec:calibration_runs}). Counts of triple coincidence from the Scintillation Counters (SCs) provided the total number of incident protons.


\section{Results}
\label{sec:results}

\subsection{Calibration Runs}
\label{subsec:calibration_runs}

\begin{figure}[!ht]
  \centering
  \includegraphics[width=0.48\textwidth]{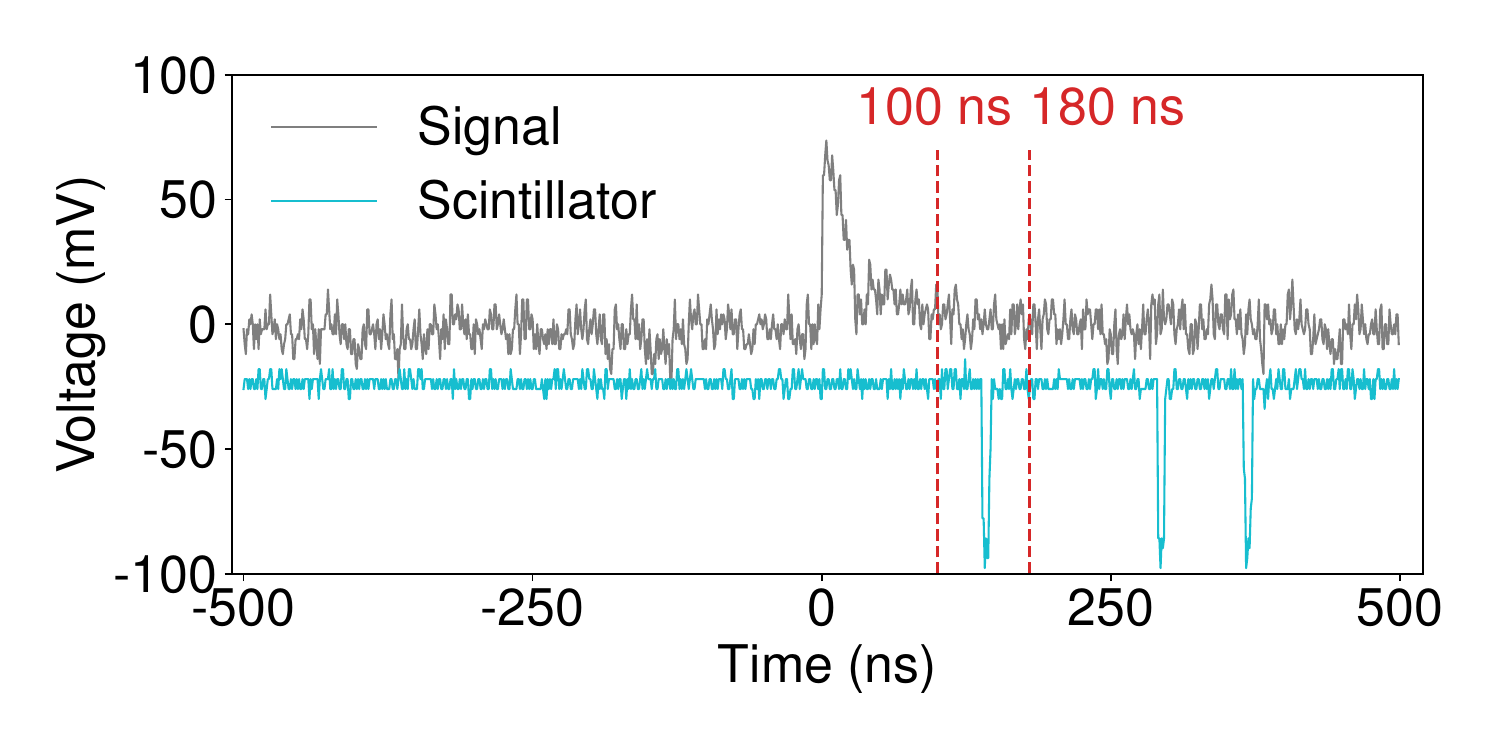}
  \caption{\label{fig:protonSignal}An example waveform that was defined as the proton signal. The grey waveform is the signal from the nanowire fire, and the light blue waveform is the one from the scintillators. Note that the scintillator signals within the timing window were only relevant to define the proton signal at the nanowire to avoid misidentification of the background count as the signal. }
\end{figure}

During the calibration runs, the 400 nm device was tested (a) to identify the waveforms for the hit candidates, (b) to determine the timing window between the SCs and the SNSPD signal, (c) to pinpoint the optimal beam position. For the calibration runs with the \SI{400}{\nm} device ($I_c$=\SI{25.2}{\uA}), the beam current ($\sim$\SI{20}{\nA}) was chosen to maximize detection efficiency while still suppressing background counts to ensure that most hits are from protons. The typical waveform from the nanowire hotspot is shown at Fig.~\ref{fig:protonSignal}. 
The measurement was done by self-triggering on the nanowire waveform within the \SI{4.2}{\s} spill, where the SCs were concurrently recorded by the same oscilloscopes. 

\begin{figure}[!ht]
  \centering
  \includegraphics[width=0.48\textwidth]{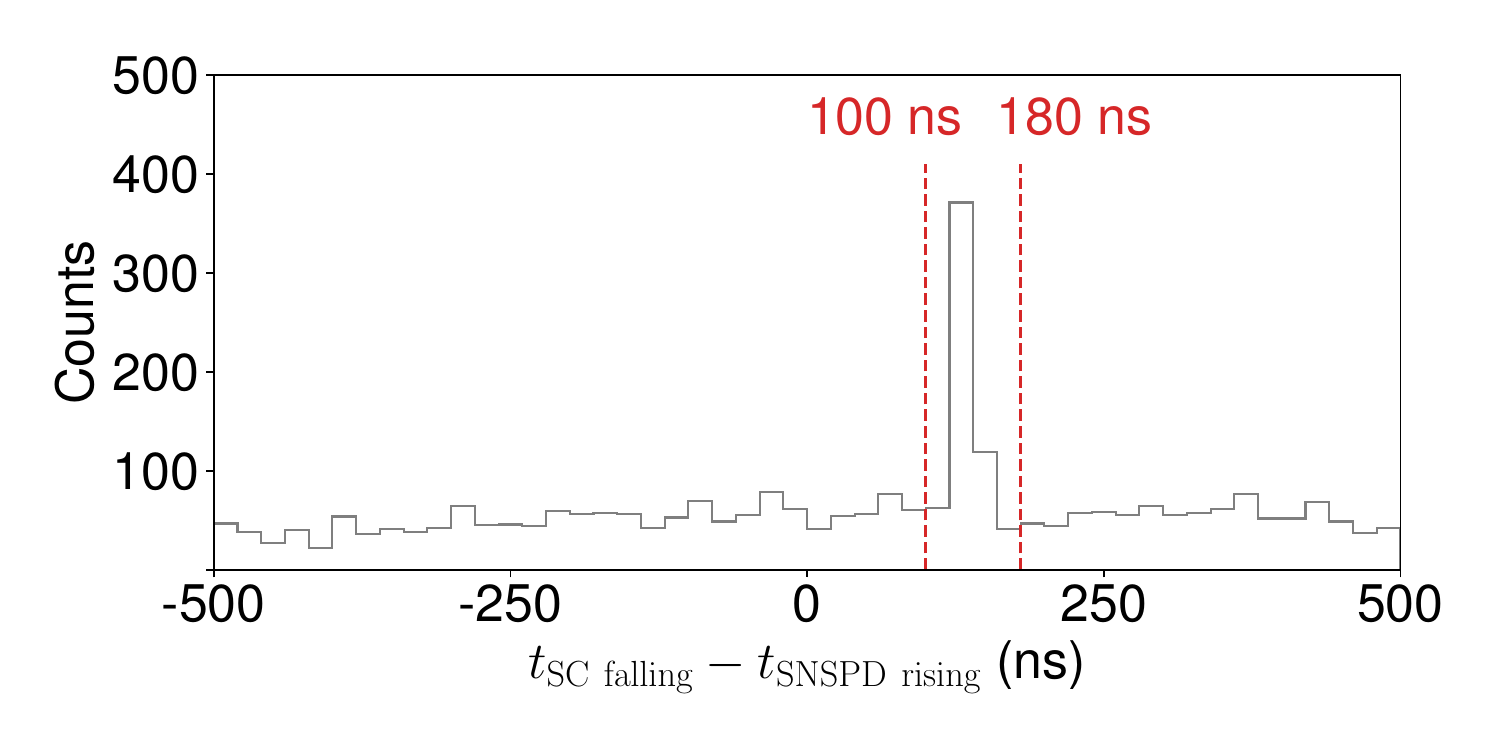}
  \caption{\label{fig:timing} The survey of timing differences between the detector fires and the scintillator signals. The timing window associated with hits is within the \SI{100}{\ns} and \SI{180}{\ns} after the detector fire.}
\end{figure}

The timing signal was defined as $t_{\text{SC}} - t_{\text{SNSPD}}$, the difference between the edges of the SNSPDs and the SC three-fold coincidence signal.  Fig.~\ref{fig:timing} shows a clear coincidence peak in the timing signal histogram. The coincidence timing window for the SC signal was conservatively set to be between \SI{100}{\ns} and \SI{180}{\ns} range.
From this stage, the proton hit was defined as the characteristic nanowire waveform triggers followed by the SC 100--\SI{180}{\ns} after. Estimated background to signal level during the calibration run was $(2.08\pm0.89)$\% (See \ref{subsec:biasscan} for the definition of background).

\begin{figure}[!ht]
  \centering
  \includegraphics[width=0.48\textwidth]{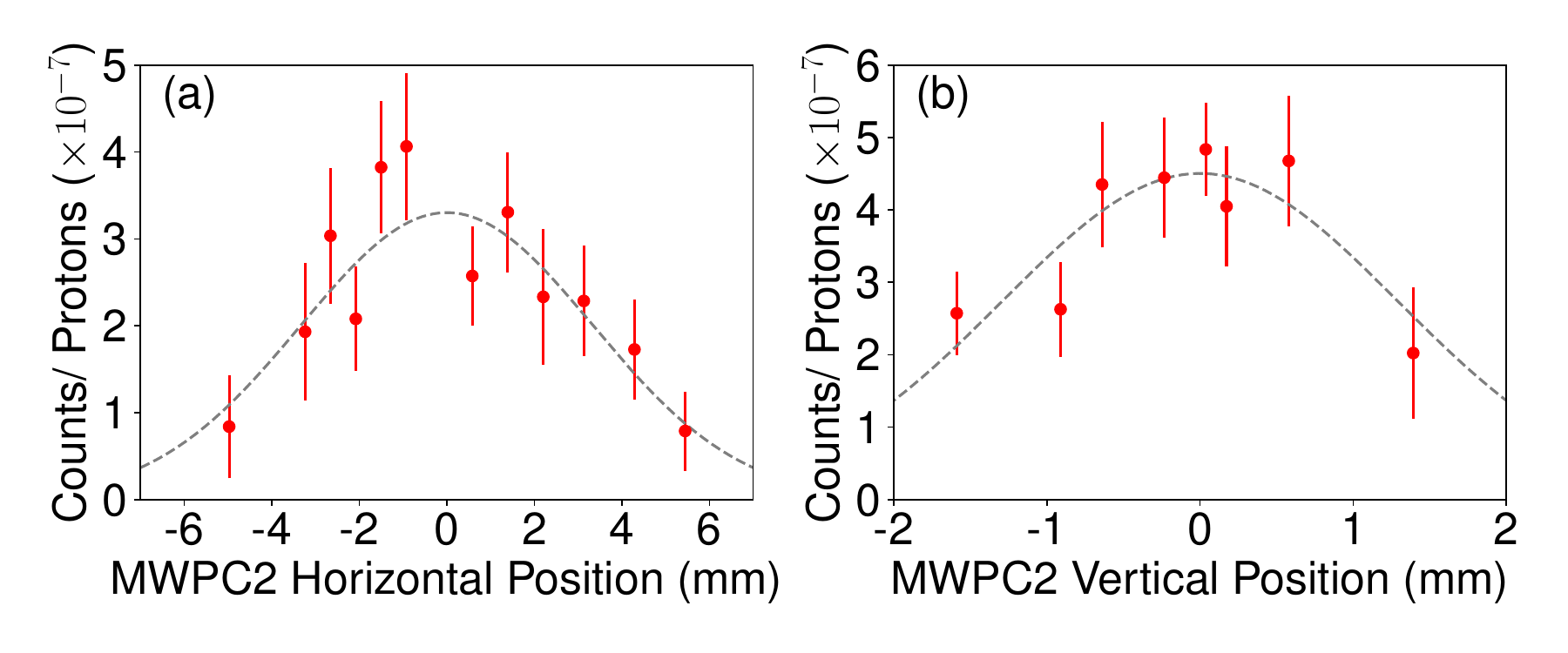}
  \caption{\label{fig:beamPos} The plots showing the normalized counts with respect to (a) horizontal and (b) vertical beam positions at MWPC2. The center position was determined by fitting the data (red points) by the Gaussian profile. The beam magnet currents were adjusted to match the MWPC2 center position with the reference determined at these plots for other runs.}
\end{figure}

Several beam positions were scanned during this run period to find the optimal beam position using the MWPCs as mentioned in Section \ref{sec:experiment}. Only few wires of the MWPC3, the closest MWPC to the SNSPD operated properly during data taking. Therefore, the MWPC2 was used as the position reference. As the cryostat was placed on a fixed table, the beamline focusing magnet currents were adjusted to steer the beam position as measured by the MWPC. The beam position scan was performed to maximize the proton hit counts normalized to the incident protons (Fig.~\ref{fig:beamPos}). Using the optimal beam position provides a consistent beam profile and thus flux seen by the superconducting sensors. This allowed for a reliable relative efficiency between sensors with different wire widths, but is limited in the extraction of the absolute efficiency due to a relatively large systematic normalization uncertainty.

\subsection{Production Runs}
\label{subsec:biasscan}

After calibration, bias scans were performed to characterize the relative efficiencies of devices with different wire widths as a function of bias current, $I_b$. The bias scan measures the proton count rate with respect to the bias currents. The tested chips included the nanowire of 300, 400, 600, and 800 nm widths. Likewise to Section \ref{subsec:calibration_runs}, proton counts were normalized to the number of incident protons, which was estimated from the total SC counts. The bias current $I_b$ was normalized to the critical currents of each device to collect the measurement results with the different widths. The critical current $I_c$ was determined as the current where electronic oscillation starts to happen. The determined $I_c$'s were \SI{12.5}{\uA} for \SI{300}{\nm}, \SI{25.2}{\uA} for \SI{400}{\nm}, \SI{48}{\uA} for \SI{600}{\nm}, \SI{55}{\uA} for \SI{800}{\nm} devices. 

\begin{figure}[!ht]
  \centering
  \includegraphics[width=0.4\textwidth]{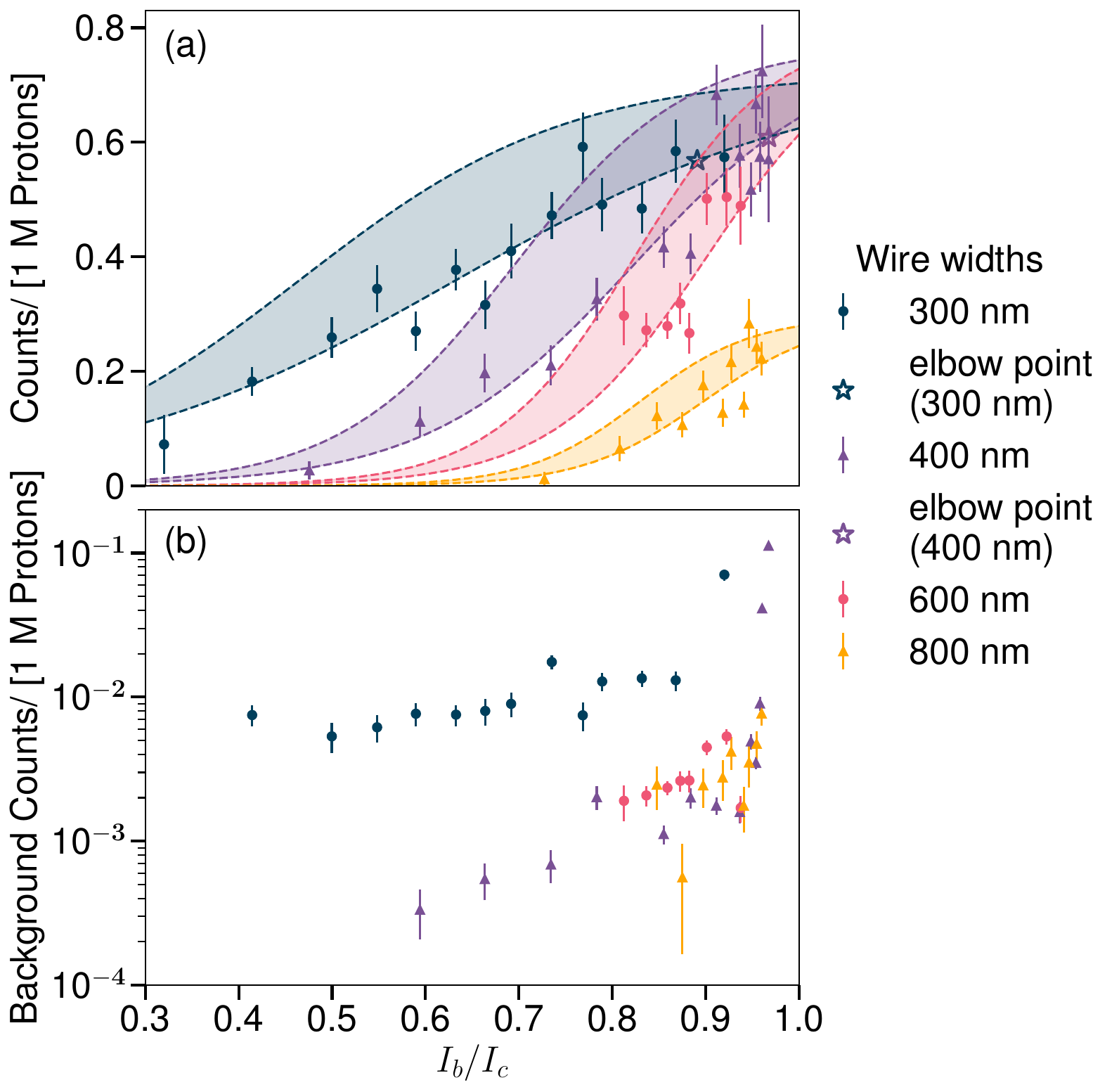}
  \caption{\label{fig:biasScan_count_and_background} The plots showing (a) normalized counts and (b) normalized background counts estimate with respect to the normalized bias current $I_b/I_c$ of the \SI{300}{\nm} (dark blue circle), \SI{400}{\nm} (purple triangle), \SI{600}{\nm} (pink circle), \SI{800}{\nm} (orange triangle) devices. Curves fit results for normalized counts with logistic functions are present, and elbow positions are visualized for \SI{300}{\nm} and \SI{400}{\nm} devices. Uncertainties on $I_b/I_c$ from underestimation of $I_c$ is visualized as bands at the top pane (a).}
\end{figure}

The collected measurement results are shown at Fig.~\ref{fig:biasScan_count_and_background}. Fig.~\ref{fig:biasScan_count_and_background}-a shows the normalized counts which generally increases with the bias current. Fig.~\ref{fig:biasScan_count_and_background}-b shows the normalized background counts estimate within 80 ns coincidence timing window. To estimate the background within the timing window, the number of detector fires outside the timing coincidence was counted, and corrected by the ratio of the time intervals of inside and outside the coincidence window.
For all devices, it is evident that the background counts start to grow exponentially around at $I_b/I_c>0.8$. At lower $I_b/I_c$, the background counts are dominated by electromagnetic interference induced detector counts, which is flat, independent of $I_b$. The interference causes an underestimation of $I_c$, and from an independent measurement it was confirmed that $I_c$ at \SI{2.82}{K} was \SI{4.5}{\uA} higher than what was determined at the FTBF. Uncertainties on $I_b/I_c$ from this effect are present on the fit curves at Fig.~\ref{fig:biasScan_count_and_background}-a as bands. It is clear that the \SI{300}{\nm} and \SI{400}{\nm} devices' relative efficiencies increase with $I_b/I_c$ at low $I_b/I_c$ region and saturate at the elbow point, where the slope change is the most visible. The elbow point can be formally defined as the point, where the third derivative of fitting curve is zero. When the data points are fitted by a logistic function, $f(I_b/I_c) = a/\left(1+\exp\left(-(I_b/I_c-b)/c\right)\right)$, the top elbow point is $b+ \ln(2+\sqrt{3})c \sim b+1.317c$ that are presented at Fig.~\ref{fig:biasScan_count_and_background}-a for the \SI{300}{\nm} and \SI{400}{\nm} devices.
Past this point, the internal detection efficiency of the device is saturated at close to 100\% internal efficiency, as confirmed by linear scaling of counts with optical power on similar devices used in reference~\cite{Polakovic:2019wrh} or by direct photon counting from calibrated sources as done in literature~\cite{lusche2014effect,marsili2011single}.

As the detection mechanism of the nanowire requires production of non-equilibrium particles that diffuse outwards to suppress the superconducting critical current, there exists a wire size where this formation and diffusion cannot happen at timescales characteristic to the quasi-particle recombination~\cite{engel2014detection}. The detection efficiency never saturates at the maximum values once the wire width is bigger than this size, as can be seen in the case of the two wider nanowire sensors. While more measurements are required to fully characterize the detection process and determine the dominant contributors (e.g., by acquiring more statistics to see activation thresholds like in ref.~\cite{verevkin2002detection}), an estimation of the characteristic length scale useful for practical applications can be made using simple thermal physics.

In a thin film with thickness much smaller than the superconducting magnetic penetration depth, and on the order of the superconducting coherence length, the problem can be treated as 2-dimensional, with the energy source due to incident particle being approximately point-like. If the diffusion of the initial proton-induced population of hot electrons is slow compared to the scattering processes that created this population, the initial hot spot will have a "hard" non-superconducting core with the size~\cite{nano10061198,sherman1962superconducting, brooks1956nuclear, kelsch1962observation}:

\begin{equation}
    r_{s} \approx \sqrt{\frac{Q}{e \pi c \rho\left(T_c-T_0\right)}},
    \label{eqn:detm}
\end{equation}

where $Q$ is the energy that the proton has deposited into the thin film, $c$ is the specific heat of the material, $\rho$ is the mass density, $T_c$ is its superconducting critical temperature and $T_0$ is the substrate temperature. This ``hard'' hot-spot reduces the effective cross-section of the nanowire and also proportionally reduces the critical current of the sensor. For simplicity, we assume that a detection event can happen only if the total current goes over the critical current of the sensor and there are no other contributing factors (e.g., there are no fluctuations that give rise to the non-saturated detection efficiency at intermediate current bias). The condition for a detection event to occur then becomes

\begin{equation}
    \frac{r_s}{w/2 \times (1 - I_b / I_c)} \geq 1,
    \label{eqn:detineq}
\end{equation}

where $r_s$ is defined by equation~\ref{eqn:detm}, $w$ is the wire width and, $I_b / I_c$ is the reduced current.

\begin{figure}
    \centering
    \includegraphics[width=0.48\textwidth]{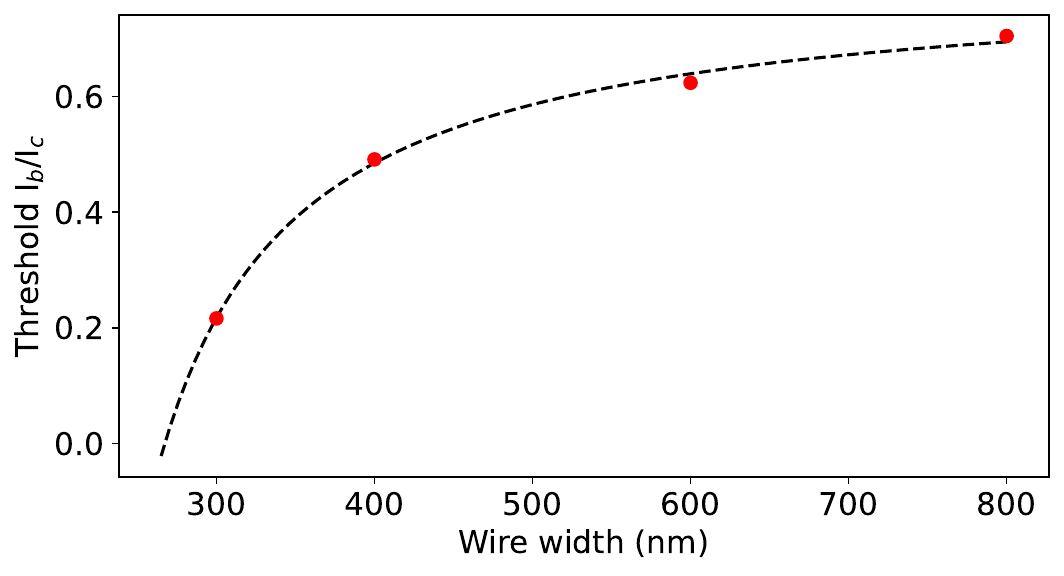}
    \caption{Reduced current at which detection events start to occur as a function of wire width. Data points are data extracted from linear fits to Fig.~\ref{fig:biasScan_count_and_background}-b, dashed line is a least-squares fit to function of equation~\ref{eqn:detineq}.}
    \label{fig:detm}
\end{figure}

By using equation~\ref{eqn:detineq} and finding points where the strict equality holds, one can extrapolate the nanowire width at which arbitrarily small bias current will cause a detection event. At this size, the detected proton deposits enough energy to create a thermal hot-spot that covers the nanowire edge-to-edge and there's no need for the avalanche process typical of the photons detection case. The extrapolated value for this ``hard'' hot-spot size in our case is $r_s$~=~\SI{134}{\nm} (Fig.~\ref{fig:detm}), which is approximately an order of magnitude bigger than the hot-spot size caused by the absorption of a photon~\cite{polyakova2019protocol} and in agreement with the estimated hot-spot size for particles with this energy~\cite{nano10061198}. While the physical validity of this simple model is a question of future work, the characteristic length scale it provides is of high importance for practical particle sensor design because it sets the optimal physical dimensions of the nanowire, which should have a width approximately smaller than double this value.

\subsection{Future Beam Tests}
\label{subsec:limitation}
To study the absolute detection efficiency, we intend to integrate pixel arrays with a tracking telescope to provide the best normalization and begin to characterize particle tracking capabilities. Furthermore, precision measurements are required to resolve geometry induced differences -- overall nanowire pixel size, wire width, wire spacing, and pixel pitch can affect the detection efficiency. 
The present results clearly indicate pixels with smaller wire widths provide a broader range of operational bias currents, however, for high energy proton detection, it is not yet clear to what extent the wire spacing impacts the efficiency of a pixel with fixed wire width. Similarly, we intend to investigate the impact of pixel spacing for arrays of devices detecting high energy particles. To address these questions, we plan to leverage one of the silicon tracking telescopes at FTBF \cite{Kwan:2014gaa} for sub-\SI{10}{\um} position resolution.

Beyond the detection performance and spatial resolutions, we intend to investigate the timing performance with 2 detector planes at FTBF's high tracking rate area, located upstream of the beam collimator for MT6. In order to reduce the timing jitter associated with the readout electronics at room temperature, we will use a newly developed cyro-CMOS ASIC~\cite{braga_cpad}. This puts the TDC measurement near the sensor which reduces the thermal load from cabling while also increasing the number of readout channels. The overall goal is to investigate the intrinsic nanowire time jitter associated with high energy particle detection, which is expected to be on the order of a few \si{\ps}~\cite{Korzh:2018oqv}.

\section{Summary}
\label{sec:summary}
We have demonstrated the direct detection of 120 GeV protons with SNSPDs with different wire widths. 
This first test highlights the viability of the SNSPDs as particle detectors at the accelerator-based nuclear and particle physics experiments. 
The relative detection rates indicate those wires with size smaller than \SI{400}{\nm} are efficient for high energy proton sensing, with an ideal wire size for this application of about \SI{250}{\nm}.
Previous work has already shown the successful SNSPD operation at magnetic field strengths typical of accelerator superconducting magnets~\cite{Polakovic:2019wrh}. 
An ongoing effort aims at developing a full scale hybrid cryogenic readout system using superconducting electronics \cite{draher2023design, 10.1063/5.0144686, 10.1063/5.0144685, castellani2023nanocryotron} and cryo-CMOS ASICs \cite{braga_cpad}. Furthermore, a radiation hardness characterization is also planned in the near future~\cite{eic-generic-rnd}.
Together with previous and ongoing R\&D, the present results open the door for the use of SNSPDs in numerous high-impact particle detection applications at accelerator facilities including the EIC.









\section*{Acknowledgements}
We would like to thank Fermilab Test Beam Facility staff and engineers for their technical support and infrastructure accommodations essential for these measurements.
%
%
This material is based upon work supported by the U.S. Department of Energy, 
Office of Science, Office of Nuclear Physics, under contract number 
DE-AC02-06CH11357 and the Office of Science under the Microelectronics Co-Design Research Project “Hybrid Cryogenic Detector Architectures for Sensing and Edge Computing enabled by new Fabrication Processes” (LAB 21-2491). A portion of this work was conducted at the Center for Nanoscale Materials, a U.S. Department of Energy, Office of Science (DOE-OS) user facility.

\bibliography{references}
\bibliographystyle{elsarticle-num}

\end{document}